\documentclass{cta-author}   

\usepackage[T1]{fontenc}
\usepackage[utf8]{luainputenc}
\usepackage[english]{babel}
\usepackage{array}
\usepackage{float}
\usepackage{epsfig}         
\usepackage{graphics}       
\usepackage{graphicx}
\usepackage{subfigure}
\usepackage{amsmath}
\usepackage{amssymb}
\usepackage{amstext}
\usepackage{amsfonts}
\usepackage{amssymb}
\usepackage{amsthm}
\usepackage{marvosym}

\usepackage[outdir=./]{epstopdf}

\usepackage{verbatim}
\usepackage{cuted}
\usepackage{lipsum}
\usepackage{pdfpages}       
\usepackage{multicol}       
\usepackage{bm}
\usepackage{lineno}

\newtheorem{problem}{Problem}
{}
{}
{}

\begin{document}

\title{Gain Scheduling LPV Control Scheme for the Autonomous Guidance Problem using a Dynamic Modelling Approach}

\author{\au{E.Alcala$^{1}$} 
\au{V. Puig$^1$}            
\au{J. Quevedo$^1$}         
\au{T. Escobet$^1$}         
}

\address{\add{1}{Advanced Control Systems Group, Automatic Control Department, Universitat Polit\`{e}cnica de Catalunya (UPC), Rambla Sant Nebridi, 10, 08222, Terrassa, Spain}
\add{}{{\fontsize{11pt}{13.2pt}\selectfont This paper is a preprint of a paper submitted to IET Control Theory \& Applications. If accepted, copy of record will be available at the IET Digital Library.}}
\email{eugenio.alcala@upc.edu}}


\begin{abstract}

This work proposes a solution for the longitudinal and lateral control problem of urban autonomous vehicles using a gain scheduling LPV control approach.
Using the kinematic and dynamic vehicle models, a linear parameter varying (LPV) representation is adopted and a cascade control methodology is proposed for controlling both vehicle behaviours.
In particular, for the control design, the use of both models separately lead to solve two LPV LMI-LQR problems.
Furthermore, to achieve the desired levels of performance, an approach based on cascade design of the the kinematic and dynamic controllers has been proposed.
This cascade control scheme is based on the idea that the dynamic closed loop behaviour is designed to be faster than the kinematic closed loop one.
The obtained gain scheduling LPV control approach, jointly with a trajectory generation module, has presented suitable results in a simulated city driving scenario.

\end{abstract}

\maketitle

\section{Introduction}

The European Parliament Research Service (EPRS) considers {\em autonomous driving} as one of the top ten technologies that will change citizen's life the most \cite{EuroParl}. This comes with no surprise given the clear benefits that one can foresee, in particular:
(1) achievement of almost zero traffic accidents by taking humans out of the driving task;
(2) inclusion of citizens with low physical mobility by the introduction of door-to-door transportation services;
(3) reduction of congestion by route sharing (passengers and goods) and a centralized mobility intelligence;
(4) decrease of energy consumption and pollution by relying on electric vehicles with a smarter vehicle control. \\

\noindent Recently, industrialized countries carry out a technological race towards autonomous driving. Research institutions and powerful companies of the automotive, mobility, and software sectors are accelerating their achievements by a great investment in human and technology resources. In spite of the gigantic difficulties of reaching full autonomy at all times and in all the places, the recent advances in hardware (sensors, embedded super-computers, etc.), software (artificial intelligence, planning and control, telecommunications, etc.), laws, and potential user acceptance seem to indicate that reaching autonomous driving is just a matter of time. \\

\noindent Nowadays, half the world’s population live in cities, and the World Health Organization predicts that by 2050 this proportion will  increase to $66\%$ \cite{WorldUrmanizationProspects}, straining city traffic more and more.
City scenarios, then, are of high relevance and present routes, speeds, traffic signs, infrastructure elements, and surroundings that are much more difficult to understand than on highways or segregated lanes, not to mention the need to take care of pedestrians and cyclists. \\

\noindent Overall, the process will be an incremental approach, with a long period of coexistence between human and artificial drivers. For that purpose, the Society of Automotive Engineers (SAE) defines 5 progressive levels of automation \cite{SAE}, from driver assistance (Level 1) to full automation (Level 5), which is not expected before 2025.
From Level 2 to Level 5 the vehicle takes full control of the accelerating, braking, and steering tasks. This means that an automatic control software will tackle the velocity and position control while handling a suitable dynamic behaviour. \\

\noindent The automatic control is one of the most important tasks inside the autonomous driving system. It is in charge of moving the vehicle between two points as well as generating smooth control actions for achieving a comfortable journey.
This is a complex task since, in addition to implement control software, dealing with real data management through sensors and actuators is needed.
First, it is required to measure vehicle variables and understanding the environment with sensors (GPS, IMU, encoders, cameras, LIDAR, etc) located around the vehicle. Then, commanding proper signals to actuators (steering motor, electric engine and brake system) to perform the motion of the vehicle.
\noindent This problem is normally defined by three general aspects: the type of control (lateral, longitudinal or both), the complexity of the model to be controlled (kinematic, linear dynamic, non-linear simplified dynamic or non-linear dynamic) and the control strategy to be used. \\

\noindent Until now, different control problems have been treated such as the longitudinal control, the lateral control and the mixed one, that includes both cases.
The goal in the longitudinal control task is to maintain the linear velocity of the vehicle around a given velocity set point that is also known as cruise control. At this point, the driver is released of the accelerating and braking tasks, being the autonomous system the responsible.
This case is included in the level 1 of automation defined by the Society of Automotive Engineers (SAE) \cite{SAE}.
On the other hand, the lateral control is in charge of controlling the yaw movement of the vehicle. To do so, the controller acts over the angle of the front wheels.
This case is the opposite one of the longitudinal control task. In this case, the driver only controls the acceleration and brakes, being the automatic controller in charge of turning.
The last control problem is the mixed one. In this case, the vehicle governs the complete 2D motion, i.e., full control of the accelerating, braking, and steering tasks and rises to the levels 2-5 of automation. \\

\noindent In turn, every one of these control problem designs rely on a vehicle model.
On one hand, kinematic models are a function of vehicle geometry. On the other hand, dynamic ones rely on physical models to describe the interaction between the vehicle and the road.
Section \ref{sec:Modelado} goes in deep in this topic. \\

\noindent The third aspect that defines the autonomous guidance problem is the choice of the control strategy. This selection is often being nested with the selection of the vehicle model, i.e., a linear model will require a linear technique while a non-linear one will need a non-linear strategy.
Nowadays, there exist several control strategies and families each one with different advantages to the application to the autonomous guidance problem.
Some of the most relevant strategies in the autonomous driving field are: Proportional-Integral-Derivative (PID), $H_{\infty}$, Fuzzy logic control, Sliding Mode Control (SMC), Lyapunov-based control, Linear Parameter Varying (LPV), Takagi-Sugeno (T-S), Linear Quadratic Regulator (LQR) and Model Predictive Control (MPC), see Table \ref{table:ControlTechniquesClassification}. \\

\noindent In order to address the real autonomous guidance situation, the longitudinal and lateral vehicle behaviours are the interesting ones to be solved, that is, the mixed control problem. \\

\noindent At this point, and taken into account only the mixed control problem, a classification of the different control strategies according to the type of model has been made.
This labelling will serve in order to illustrate what kind of solutions are being more used for solving the more complex autonomous guidance problem.
Table \ref{table:ControlTechniquesClassification} shows such a classification with corresponding references.

\begin{table}[htbp]
    \caption{Classification of control techniques according to the type of model.}
    \begin{tabular}{|l|l|l|}
    \hline
     & \multicolumn{2}{|c|}{Mixed Control Problem} \\
    \hline
    \multicolumn{1}{|c|}{Control Strategy} & \multicolumn{1}{|c|}{Kinematic Model} & \multicolumn{1}{|c|}{Dynamic Model}  \\
    \hline \hline
    $H_{\infty}$ & \cite{H_infinity} &  \\
    SMC/SMC Adaptive & \cite{alcala2016comparison} & \\
    Lyapunov  & \cite{indiveri1999kinematic}, \cite{alcala2016comparison}, \cite{blavzivc2010takagi}, \cite{LyapunovBasedControl} &  \\
    LPV &  & \cite{nemeth2016lpv} \\
    T-S & \cite{blavzivc2010takagi} & \\
    MPC & \cite{LinearMPC},\cite{gonzalez2011robust} & \cite{LinearMPC} \\
    Non-linear MPC & \cite{farrokhsiar2013integrated} & \cite{NLMPC}, \cite{JournalMPC} \\
    \hline
    \end{tabular}
    \label{table:ControlTechniquesClassification}
\end{table}

\noindent Some of the previously enumerated control strategies (such as Proportional-Integral-Derivative (PID), Fuzzy logic control or Linear Quadratic Regulator (LQR)) do not appear in the table due to they mostly have been applied to solve the lateral control problem. \\

\noindent The paper is structured as follows: Section \ref{sec:Modelado} presents and describes the different types of vehicle models used for control purposes. In Section \ref{sec:LPV_modelling}, the LPV modelling is developed. Section \ref{sec:vehicle_controllers} presents the feedback control design using the gain scheduling LPV approach. Finally, Section \ref{sec:results} shows the results and Section \ref{sec:conclusions} presents the conclusions of the work.


\section{Kinematic and Dynamic Vehicle Models}
\label{sec:Modelado}
\begin{figure}[h]
  \centering
  \includegraphics[width=86mm]{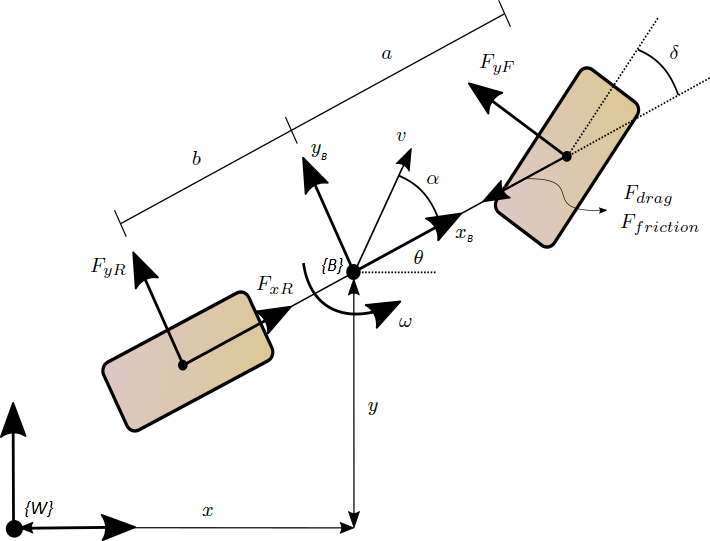}
  \caption{ Two-wheels bicycle model used for control purposes. $\{ W \}$ frame represents the global inertial frame and $\{ B \}$ is the body frame located in the centre of gravity of the vehicle.}
  \label{fig:vehicle_model}
\end{figure}

This section describes the models which will be used later for developing the automatic control strategies.
The behaviour of a mobile object can be described by using equations that represent the dynamic and kinematic performances.
Unlike common mobile robots, urban autonomous vehicles are systems with larger mass and operating at a higher velocity. Hence, the use of dynamic models becomes indispensable.
On one hand, in dynamic models the sum of forces existing over the vehicle are taken into account for computing the vehicle acceleration.
The motion is generated by applying forces over the driven wheels and mass, inertial and tire parameters are considered.
On the other hand, kinematic model is based on the velocity vector movement in order to compute longitudinal and lateral velocities referenced to a global inertial frame.
External forces are not considered in this case.
Note that, for both models, the two-wheels bicycle model has been considered as the one depicted in Fig. \ref{fig:vehicle_model}.
It is interesting to specify that the two-wheels model employed does not consider consider $roll$, $pitch$ and $z$ motion, only $yaw$, $x$ and $y$ movements. \\

\noindent In this work, both models are presented and used in a decoupled way. It means that, both model behaviours will be controlled in a decoupled way by using two control loops.
Table \ref{table:parameters} presents the characteristic vehicle parameters used in the models.

\begin{table}[htbp]
\caption{Kinematic and dynamic model parameters.}
\label{table:parameters}
\begin{tabular}{|l|l|l|}
\hline
 Parameter & Description & Value \\
\hline \hline
$a$     & Distance from CoG to front axle   & 0.758 $m$   \\ \hline
$b$     & Distance from CoG to rear axle    & 1.036 $m$   \\ \hline
$M$     & Vehicle mass                      & 683 kg    \\ \hline
$I$     & Vehicle yaw inertia               & 560.94 $kg$ $m^2$  \\ \hline
$C_d$   & Drag coefficient                  & 0.36       \\ \hline
$Ar$     & Vehicle frontal area              & 1.91 $m^2$ \\ \hline
$\rho$  & Air density at 25$^{\circ}$C      & 1.184 $\frac{kg}{m^3}$  \\ \hline
$\mu$   & Friction coefficient              & 0.09  \\ \hline
$C_x$   & Tire stiffness coefficient        & 25000 $\frac{N}{rad}$ \\ \hline
\end{tabular}
\end{table}

\subsection{Kinematic model}
Kinematic based model is widely used due to its low parameter dependency. It assumes null skidding and considers lateral force to be so small that can be neglected. Basically, it is a geometric manner to compute vehicle position and orientation as of linear and angular velocities.
The kinematic equations are introduced below:
\begin{equation}
    \begin{cases}
        \dot x = v \cos(\theta)  \\
        \dot y = v \sin(\theta) \\
        \dot \theta = \omega
    \end{cases}
    \label{eq:kinematicModel_real}
\end{equation}

\noindent where:
\begin{itemize}
    \item $x$, $y$ and $\theta$ represent the current position and orientation of the vehicle in meters ($m$) and radians ($rad$), respectively, with respect to the inertial frame $\{$w$\}$.
    \item $v$ is the linear velocity in $\frac{m}{s}$.
    \item $\omega$ represents the vehicle angular velocity in $\frac{rad}{s}$.
\end{itemize}

\subsection{Dynamic model}

The dynamical behaviour of a vehicle is generally complicated to represent it in a detailed manner.
In practical applications, normally simplified models are used. In this case, the obtained model is based on the second Newton's law. The streamlined dynamical model of the road vehicle can be written as:

\begin{equation}
    \label{eq:dynamic_model}
    \begin{aligned}
        & \dot v = \frac{F_{xR} \cos(\alpha) + F_{yF} \sin(\alpha - \delta) + F_{yR} \sin(\alpha) - F_{df}}{M}  \\
        & \dot \alpha =  \frac{-F_{xR} \sin(\alpha) + F_{yF} \cos(\alpha - \delta) + F_{yR} \cos(\alpha)}{M v} - \omega \\
        & \dot \omega = \frac{ F_{yF} a \cos(\delta) - F_{yR} b}{I}
    \end{aligned}
\end{equation}

\begin{equation}
    F_{yF} = C_x \left( \delta - \alpha - \frac{a \omega}{v} \right)
\end{equation}
\begin{equation}
    F_{yR} = C_x \left( -\alpha + \frac{b \omega}{v}\right)
\end{equation}
\begin{equation} \label{eq:last_model_equation}
    F_{df} = F_{drag} + F_{friction} = \frac{1}{2} C_d \rho Ar v^2 + \mu M g
\end{equation}

\noindent where:
\begin{itemize}
    \item $\alpha$ represents the vehicle slip angle ($rad$).
    \item $\delta$ is the steering angle and one of the inputs of the system ($rad$).
    \item $F_{xR}$ is the longitudinal rear force and the other input of the system ($N$).
    \item $F_{yR}$ represents the lateral rear force that appears when steering ($N$).
    \item $F_{yF}$ is the lateral front force which appears also with the angular motion ($N$).
    \item $F_{drag}$represents drag force that opposes to the forward movement ($N$).
    \item $F_{friction}$ is the friction force that also opposes to the longitudinal vehicle movement ($N$).
\end{itemize}

\noindent Note that instead of employing the states $x$ and $y$, a new representation has been adopted by using the polar representation and considers the variables $v$ and $\alpha$. These variables can be seen in Figure \ref{fig:vehicle_model}.
Observe that the dynamic model variables are referred to the vehicle body frame $\{ B \}$ while the kinematic set of variables refers to the global fixed coordinate system $\{ W \}$ in order to represent the trajectory from a relative point of view.


\section{LPV Control Oriented Model}
\label{sec:LPV_modelling}

The LPV control technique needs of a linear-like representation of the non-linear model to be controlled. Hence, the LPV modelling task is presented in this section.
This method consists on embedding the model non-linearities inside model parameters that depend on some variables, called scheduling variables, that vary in a known bounded interval.
In the last section kinematic and dynamic non-linear models were presented. Here, a LPV representation for each one is adopted.
For the kinematic LPV modelling task, a reference model has been built previously.
Note that two decoupled LPV models have been obtained in order to control the kinematic and dynamic parts of the vehicle separately.


\subsection{Kinematic LPV modelling}

To obtain the kinematic LPV model, a reference model has been developed.
This model is defined as the difference between real measurements ($x, y$ and $\theta$) and desired values ($x_d,y_d$ and $\theta_d$).
However, this set of errors are expressed with respect to the inertial global frame $\{ W \}$ (see Figure \ref{fig:vehicle_model}).
For control purposes is suitable to express the errors with respect to the vehicle, such that the lateral error is always measured in the lateral axis of the vehicle. Thus, a rotation over the road orthogonal axis is considered to represent the errors in the body vehicle frame $\{ B \}$:
\begin{equation}
    \label{eq:error_model}
    \left[\begin{array}{c}
     x_{e} \\
     y_{e} \\
    \theta_{e} \\
    \end{array}\right]=
    \left[\begin{array}{ccc}
     \cos (\theta) & \sin (\theta) & 0 \\
     -\sin (\theta) & \cos (\theta) & 0 \\
     0 & 0 & 1 \\
    \end{array}\right]
    \left[\begin{array}{c}
     x_{d} - x  \\
     y_{d} - y \\
     \theta_{d} - \theta \\
    \end{array}\right]
\end{equation}

\noindent where subindexes $d$ and $e$ represent the desired and error values, respectively.
To develop the error model is needed to take into account the rear wheels non-holonomic constraint of the form:
\begin{equation}
    \label{eq:nonholonomic}
    \dot x \sin (\theta) = \dot y \cos (\theta)
\end{equation}

\noindent Hence, computing the time derivative of (\ref{eq:error_model}) and using (\ref{eq:kinematicModel_real}), (\ref{eq:nonholonomic}) and some trigonometric identity, we obtain the following open-loop error system:
\begin{equation}
    \label{eq:kinematic_error_model}
    \begin{aligned}
        & \dot x_{e} = \omega y_{e} + v_d \cos \theta_e - v   \\
        & \dot y_{e} =  -\omega x_{e} + v_d \sin \theta_e \\
        & \dot \theta_{e} = \omega_d - \omega
    \end{aligned}
\end{equation}

\noindent Details about the  development of \eqref{eq:kinematic_error_model} can be found in Chapter 1 of \cite{LyapunovBasedControl}.
At this point, denoting the state, control and output vectors, respectively, as
\begin{equation}
    \label{eq:kinematic_state_space_vectors}
    \bm{x} =
    \left[\begin{array}{c}
        x_e \\
        y_e  \\
        \theta_e  \\
    \end{array}\right] , \
    \bm{u} =
	\left[\begin{array}{c}
        v \\
        \omega \\
    \end{array}\right] , \
    \bm{y} =
    \left[\begin{array}{c}
        x_e \\
        y_e  \\
        \theta_e  \\
    \end{array}\right]
\end{equation}

\noindent we can obtain the LPV representation for the kinematic dynamics \eqref{eq:kinematic_error_model}. Then, considering $\omega$, $v_{d}$, $\theta_e$ $\in \mathbb{R}$ as the kinematic scheduling variables, the LPV form becomes:
\begin{subequations}
    \label{eq:kinematic_LPV_matrices}
    \begin{equation}
        \begin{cases}
            \bm{\dot x} = \bm{A}(\omega, v_d, \theta_e) \bm{x} + \bm{B} \bm{u} - \bm{B} \bm{r} \\
            \bm{y} = \bm{C} \bm{x} \\
        \end{cases}
    \end{equation}

    \noindent where
    \begin{equation}
         \bm{A} \left( \omega, v_d, \theta_e \right) =
    	\left[\begin{array}{ccc}
            0 & \omega & 0 \\
            -\omega & 0 & v_d \frac{\sin \theta_e}{\theta_e} \\
            0 & 0 & 0 \\
        \end{array}\right]
    \end{equation}
    \begin{equation}
        \bm{B} =
        \left[\begin{array}{cc}
            -1 & 0 \\
             0 & 0 \\
             0 & -1 \\
        \end{array}\right] \ , \
        \bm{C} =
        \left[\begin{array}{ccc}
             1 & 0 & 0 \\
             0 & 1 & 0 \\
             0 & 0 & 1 \\
        \end{array}\right]
    \end{equation}
    \begin{equation}
        \bm{r} =
        \left[\begin{array}{c}
             v_d \cos \theta_e \\
             \omega_d \\
        \end{array}\right]
    \end{equation}
\end{subequations}

\noindent For the control design purpose, the reference vector $\bm{r}$ will not be taken into account, only $\bm{A}$ and $\bm{B}$. This vector will be added directly to the control law.

\subsection{Dynamic LPV modelling}
\label{sec:Dynamic_LPV}

The dynamic model is quite more complex than the presented kinematic one. Thus, the development of a LPV model is more involved and therefore, it is presented in progressive steps. \\

\noindent Denoting the state, control and output vectors, respectively, as
\begin{equation}
    \label{eq:dynamic_state_space_vectors}
    \bm{x} =
    \left[\begin{array}{c}
        v \\
        \alpha  \\
        \omega  \\
    \end{array}\right] , \
    \bm{u} =
	\left[\begin{array}{c}
        F_{xR}\\
        \delta \\
    \end{array}\right] , \
    \bm{y} =
    \left[\begin{array}{c}
        v \\
        \alpha  \\
        \omega  \\
    \end{array}\right]
\end{equation}

\noindent and considering the slip angle ($\alpha$) to remain very small in comparison with $\delta$ variable, the state space model for the dynamic representation \eqref{eq:dynamic_model} can be obtained as:

\begin{subequations}
\label{eq:dynamic_state_space_formulation}
    \begin{equation}
        \begin{cases}
            \bm{\dot x} = \bm{A} (\delta, v) \bm{x} + \bm{B} (\delta, v) \bm{u}\\
            \bm{y} = \bm{C} \bm{x} \\
        \end{cases}
    \end{equation}

    \noindent where:
    \begin{equation}
        \label{eq:matrices_state_space_A_D}
        \bm{A} (\delta, v) =
        \left[\begin{array}{ccc}
            -\frac{F_{df}}{M v} & \frac{C_{x} \sin(\delta)}{M} & \frac{C_{x} a \sin(\delta)}{M v} \\

            0 & \frac{C_{x} \cos(\delta) - C_{x}}{M v} &  \frac{C_{x} a \cos(\delta) - C_{x} b}{M v^2}-1 \\

            0 & \frac{C_{x} b - C_{x} a \cos(\delta)}{I} & -\frac{C_{x} b^2 + C_{x} a^2 \cos(\delta)}{I v} \\
        \end{array}\right]
    \end{equation}
    \begin{equation}
        \label{eq:matrices_state_space_B_D}
        \bm{B} (\delta, v) =
        \left[\begin{array}{cc}
            \frac{1}{M} & \frac{-C_{x} \sin(\delta)}{M} \\
            0 & \frac{-C_{x} \cos(\delta)}{M v} \\
            0 & \frac{C_{x} a \cos(\delta)}{I} \\
        \end{array}\right]
    \end{equation}
    \begin{equation}
        \label{eq:matrices_state_space_C_D}
        \bm{C} =
        \left[\begin{array}{ccc}
             1 & 0 & 0 \\
             0 & 1 & 0 \\
             0 & 0 & 1 \\
        \end{array}\right]
    \end{equation}
\end{subequations}

\noindent The last consideration is made due to $\delta$ variable will be much greater than $\alpha$ in normal driving conditions, i.e. turning and accelerating the vehicle smoothly.\\

\noindent At this point, $\bm{A}$ and $\bm{B}$ are time varying matrices. However, with the aim of avoiding the dependency on varying parameters in matrix $\bm{B}$, the system has been augmented by adding a fast dynamic filter as suggested by \cite{apkarian1995self} in the form:

\begin{equation}
\label{eq:filter}
    \bm{\dot x}_{f} = \bm{A}_{f} \bm{x}_{f} + \bm{B}_{f} \bm{u}_{f}
\end{equation}
\begin{equation*}
    \label{eq:Apkarian_filter}
    \left[\begin{array}{c}
        \dot F_{xR} \\
        \dot \delta
    \end{array}\right] =
    \left[\begin{array}{cc}
        -\gamma & 0 \\
         0  & -\gamma
    \end{array}\right]
    \left[\begin{array}{c}
        F_{xR} \\
        \delta
    \end{array}\right] +
    \left[\begin{array}{cc}
        \gamma & 0 \\
         0 & \gamma
    \end{array}\right]
    \left[\begin{array}{c}
        u_{F} \\
        u_{\delta}
    \end{array}\right]
\end{equation*}

\noindent where $\gamma$ represents the filter gain, $u_F$ is the new longitudinal behaviour input and $u_\delta$ is the new lateral behaviour input.
Note that this new added states have fast dynamics and will not disturb the dynamic model \eqref{eq:dynamic_state_space_formulation}. \\

\noindent Then, the system \eqref{eq:dynamic_state_space_formulation} is transformed into a fifth order system with state and input vectors as

\begin{subequations}
\label{eq:dynamic_state_space_vectors_augmented}
    \begin{equation}
        \bm{x} =
        \left[\begin{array}{c}
            v \\
            \alpha  \\
            \omega  \\
            F_{xR} \\
            \delta\\
        \end{array}\right] , \
        \bm{u_f} =
    	\left[\begin{array}{c}
            u_{F} \\
            u_{\delta}
        \end{array}\right]
    \end{equation}

    \noindent and matrices $\bm{A}$ and $\bm{B}$ as

    \begin{equation}
        \bm{A} (\delta, v) =
    	\left[\begin{array}{ccccc}
            A_{11} & A_{12} & A_{13} & \frac{1}{M} & A_{15}\\
            0 & A_{22} & A_{23} & 0 & A_{25}\\
            0 & A_{32} & A_{33} & 0 & A_{35} \\
            0 & 0 & 0 & -\gamma & 0 \\
            0 & 0 & 0 & 0 & -\gamma \\
        \end{array}\right]
    \end{equation}
    \begin{equation}
        A_{11} = -\frac{F_{df}}{Mv} \ , \
        A_{12} = \frac{C_{x} \sin(\delta)}{M} \\
    \end{equation}
    \begin{equation}
        A_{13} = \frac{C_{x} a \sin(\delta)}{M v} \ , \
        A_{15} = -\frac{C_{x} \sin(\delta)}{M} \\
    \end{equation}
    \begin{equation}
        A_{22} = \frac{C_{x} \cos(\delta) - C_{x}}{M v} \ , \
        A_{23} = \frac{C_{x} a \cos(\delta) - C_{x} b}{M v^2}-1 \\
    \end{equation}
    \begin{equation}
        A_{32} = \frac{C_{x} b - C_{x} a \cos(\delta)}{I} \ , \
        A_{33} = -\frac{C_{x} b^2 + C_{x} a^2 \cos(\delta)}{I v} \\
    \end{equation}
    \begin{equation}
        A_{25} = \frac{-C_{x} \cos(\delta)}{M v} \ , \
        A_{35} = \frac{C_{x} a \cos(\delta)}{I} \\
    \end{equation}
    \begin{equation}
        \bm{B} =
        \left[\begin{array}{cc}
             0 & 0 \\
             0 & 0 \\
             0 & 0 \\
             \gamma & 0 \\
             0 & \gamma \\
        \end{array}\right]
    \end{equation}
\end{subequations}

\noindent However, the model still presents some features that will difficult the control design task.
One of them is that the input $\delta = 0$ has been identified as a singular point. Hence, to avoid it, a change of variable has been done by shifting the $\delta$ interval:
\begin{equation}
    \delta \in \left[ \underline{\delta}, \overline{\delta} \right] \to \sigma \in \left[ \underline{\delta}+\varepsilon, \overline{\delta}+\varepsilon \right]
\end{equation}
converting $\sigma$ into the new scheduling variable and being $\varepsilon$ a constant value greater than $\overline{\delta}$. \\

\noindent In addition to all these arrangements, it was found that the angular velocity channel lacks integral action, thus leading to a steady state error. Thus, the addition of such action through the controller is considered.
Then, a new state ($i_p$) has been added as the integral of the state $\omega$:

\begin{equation}
    \dot i_p = -\omega
\end{equation}

\noindent Therefore, taken into account these considerations and denoting the scheduling variables as $\sigma, v \in \mathbb{R}$, the vehicle dynamic LPV model is presented as:

\begin{subequations}
\label{eq:augmented_IP_LPV_dynamic}
    \begin{equation}
        \begin{cases}
            \bm{\dot x_D} = \bm{A}(\sigma, v) \bm{x_D} + \bm{B} \bm{u_f}\\
            \bm{y} = \bm{C} \bm{x_D} \\
        \end{cases}
    \end{equation}

    \noindent with state and input vectors

    \begin{equation}
    \label{eq:state_vector_LPV}
        \bm{x_D} = \left[\begin{array}{c}
            v \\
            \alpha \\
            \omega \\
            F_{xR} \\
            \sigma \\
            i_p \\
        \end{array}\right] \ , \
        \bm{u_f} = \left[\begin{array}{c}
            u_{F} \\
            u_{\delta}
        \end{array}\right]
    \end{equation}

    \noindent and matrices $\bm{A}$ and $\bm{B}$ as:

    \begin{equation}
        \bm{A} (\sigma, v) =	\left[\begin{array}{cccccc}
            A_{11} & A_{12} & A_{13} & \frac{1}{m} & A_{15} & 0 \\
            0 & A_{22} & A_{23} & 0 & A_{25} & 0\\
            0 & A_{32} & A_{33} & 0 & A_{35} & 0 \\
            0 & 0 & 0 & -\gamma & 0 & 0 \\
            0 & 0 & 0 & 0 & -\gamma & 0 \\
            0 & 0 & -1 & 0 & 0 & 0 \\
        \end{array}\right]
    \end{equation}
    \begin{equation}
        \bm{B} = \left[\begin{array}{cc}
             0 & 0 \\
             0 & 0 \\
             0 & 0 \\
             \gamma & 0 \\
             0 & \gamma \\
             0 & 0 \\
        \end{array}\right]
    \end{equation}
\end{subequations}

\noindent The model \eqref{eq:augmented_IP_LPV_dynamic} will be used for designing the dynamic state feedback control.


\section{Control Design using LPV Approach}
\label{sec:vehicle_controllers}
\begin{figure*}[htb!]
  \centering
  \includegraphics[width=160mm]{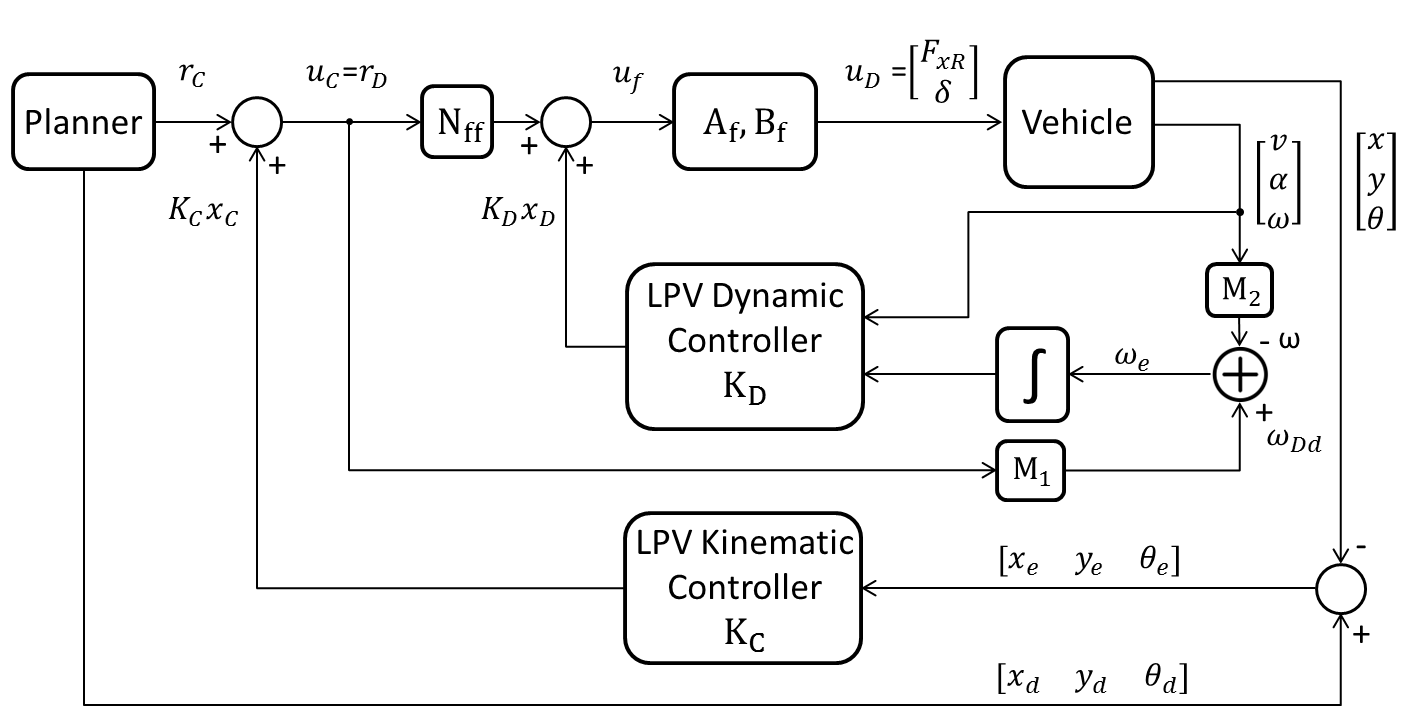}
  \caption{Complete control diagram where $\bm{M}_1 = [0 \ 1]$ and $\bm{M}_2 = [0 \ 0 \ 1]$.}
  \label{fig:control_scheme}
\end{figure*}

The automatic control strategy addresses the problem of generating an appropriate vehicle behaviour from a desired reference. In this work two cascade feedback LPV controllers are proposed for controlling appropriately the behaviour of the vehicle (see Figure \ref{fig:control_scheme}).
Furthermore, a trajectory planner \cite{GuarinoPiazziRomano} is used which is in charge of providing the correspondent position and velocities references to the kinematic controller. \\

\noindent In this approach, a cascade methodology is employed where the internal and fast loop corresponds to the dynamic control and the external one to the kinematic control.
On one hand, the kinematic control ($K_C$ in Figure \ref{fig:control_scheme}) is in charge of computing smooth control actions (linear and angular velocities) such that the vehicle is capable of achieving the required speed, position and orientation at the next local way-point.
On the other hand, the dynamic control strategy ($K_D$ in Figure \ref{fig:control_scheme}) allows the vehicle to follow the angular and linear velocity references provided by the kinematic control loop. 
To this aim, the dynamic control generates forces to the rear wheels and a steering angle signal for the front wheels.

\subsection{Description of the design method}

The LPV technique allows to use a family of systems for designing the controller.
In particular, at each operating point, the system model parameters are parameterised by a vector of scheduling variables $\bm{Sv}$.
Thus, the LPV model is denoted by:
\begin{equation}
    \label{eq:LPV_system}
    \begin{aligned}
       & \dot x (t) = \bm{A} (\bm{Sv} (t)) x(t) + \bm{B} u(t) \\
       &    y (t)   = \bm{C} x(t)
    \end{aligned}
\end{equation}

\noindent with $\bm{B}$ and $\bm{C}$ constant. 
The vector of scheduling variables is defined as $\bm{Sv} := \left[ sv_1 (t), ..., sv_{n_{sv}} (t) \right]$ being $n_{sv}$ the number of variables. Each parameter $sv_i$ is known and varies in a defined interval $sv_i (t) \in \left[ \underline{sv_i}, \overline{sv_i} \right]$.
The value and number of scheduling variables determine the size and form of a polytope with $2^{n_{sv}}$ vertices. 
Thus, the LPV system is defined by a set of $A_i$ matrices that correlate with each one of the polytope vertexes. \\

\noindent For simplicity, the time variable notation will be omitted in next equations.
Thereupon, in order to stabilize the system at each operating point for a set of arbitrary parameters $\bm{Sv}$ it is sufficient to stabilize $\bm{A}(\bm{Sv}))$ at the extremes of the parameter variation intervals, i.e. $A_i$ with $i$ = 1,...,$2^{n_{Sv}}$, using the bounding box approach \cite{apkarian1995self}.

\noindent Note also that matrix $\bm{B}$ is independent of the set of scheduling variables. Therefore, the design control problem is formulated as
\begin{problem}
\label{problem1}
    Let $\bm{A}_i$, $i$ = 1,...,$2^{n_{Sv}}$, find a set $\bm{K}_i$ of controllers that makes the closed-loop system asymptotically stable and provide an optimal performance for the family of systems $A_i$ using the state feedback control law $\bm{u}=\bm{K}\bm{x}$, where $\bm{K}$ is an interpolated matrix dependent on $\bm{K}_i$.
\end{problem}

\noindent For solving the stated problem, the linear-quadratic regulation (LQR) technique via LMI is used, as suggested in \cite{duan2013lmis} using the LMI solution for the $H_2$ problem

\begin{equation}
\label{eq:LQR-LMI-introduction}
\begin{aligned}
    & (\bm{A}_i\bm{P} + \bm{B}\bm{W}_i) + (\bm{A}_i \bm{P} + \bm{B}\bm{W}_i)^T + 2 \eta \bm{P} < 0 \\
    & \left[\begin{array}{cc}
      	    -\bm{Y} & \bm{R}^{\frac{1}{2}} \bm{W}_i \\
     	    (\bm{R}^{\frac{1}{2}} \bm{W}_i)^T & -\bm{P}
                \end{array}\right] < 0  \ , \ \ \ \ \ \ \ \ \ \ \ \ \ \ \ \ i = 1,...,2^{n_{Sv}} \\ 
    & trace(\bm{Q}^{\frac{1}{2}} \bm{P} (\bm{Q}^{\frac{1}{2}})^T )+trace(\bm{Y})<\gamma \\
    & \bm{P} \ge 0, \ \bm{Y} = \bm{Y}^{T} > 0
\end{aligned}
\end{equation}

\noindent where $\bm{Q}= \bm{Q}^T \ge 0$, $\bm{R}= \bm{R}^T > 0$, and $\gamma > 0$ are the LQR tuning variables.
Once matrices $\bm{W}_i$ and $\bm{P}$ have been obtained, each one of the vertex controllers can be computed as follows

\begin{equation}
    \label{eq:controller_obtaining}
    \bm{K}_i = \bm{W}_i \bm{P}^{-1}
\end{equation}

\noindent Note that the problem has solution if and only if there exist $\bm{P} \in \mathbb{R}^s$, $\bm{Y} \in \mathbb{R}^r$ and $\bm{W}_i \in \mathbb{R}^{r\times s}$, being $r$ the number of control actions and $s$ the number of states.
Observe also that decreasing the parameter $\gamma$ increases the performance of the control loop.
In turn, $\eta$ establishes a constraint for the decay rate. \\

\noindent Up to here the design procedure. From now on, at every control iteration, the controller will use the polytopic interpolation method suggested by \cite{apkarian1995self} based on the current operating point and computing the relative distance to vertexes in order to obtain
the set of weights $\mu_i$.
Then, the control matrix $\bm{K}$ is obtained by
\begin{equation}
    \label{eq:interpolation}
    \bm{K} (\bm{Sv}) = \sum_{i=1}^{n_{Sv}} \mu_i (\bm{Sv}) \bm{K}_i
\end{equation}

\noindent where $K_i$ represents the polytopic vertex controllers presented in Tables \ref{table:dynamic_controllers} and \ref{table:kinematic_controllers} and $\mu_i$ represents the interpolation weight that is obtained by means of a polytopic interpolation \cite{floater2006general}.

\noindent Next subsections provide details of the particular control design for the dynamic and kinematic vehicle behaviours.

\subsection{Dynamic LPV control design}
The dynamic control addresses the tracking of the linear and angular velocity references of the vehicle by applying force to the wheels and an angle to the front wheels. \\

\noindent At this point, the LPV model developed in Section \ref{sec:Dynamic_LPV} is used for solving the Problem 1. 
The chosen scheduling variables are $\sigma$ and $v$ bounded in the following intervals:
\begin{equation*}
    v \in [1, 18]  \ \frac{m}{s} \ \ \ \ \ \text{and} \ \ \ \ \sigma \in [0.0873, 0.9599]  \  rad
\end{equation*}

\begin{figure}[htb!]
  \centering
  \includegraphics[width=40mm]{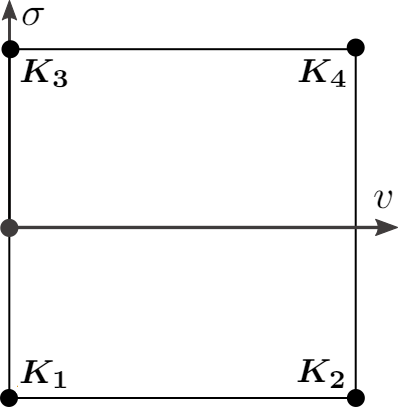}
  \caption{Dynamic polytopic region defined by two scheduling variables.}
  \label{fig:polytopic_dynamic}
\end{figure}

\noindent The proposed design matrices $\bm{Q}$ and $\bm{R}$ are presented in Table \ref{table:RMSEs} and parameter $\gamma$ is set as $0.001$. Such a Problem \ref{problem1} returns matrices $\bm{P} \in \mathbb{R}^6$, $\bm{Y} \in \mathbb{R}^2$ and $\bm{W} \in \mathbb{R}^{2 \times 6}$ that using \eqref{eq:controller_obtaining} are used for obtaining $\bm{K_i}$ (see Table \ref{table:dynamic_controllers}).
Due to dynamic LPV model has two scheduling variables the polytopic space will be a square figure as the one presented in Figure \ref{fig:polytopic_dynamic}.

\noindent The controller obtained at each control iteration follows the rule presented in \eqref{eq:interpolation} where weighting variables $\mu_i$ are computed as follows
\begin{subequations}
  \begin{align}
  \mu_1 = M_{\underline{v}} M_{\underline{\sigma}}  \ , \ \ \mu_3 = M_{\overline{v}} M_{\underline{\sigma}}  \\
  \mu_2 = M_{\underline{v}} M_{\overline{\sigma}}  \ , \ \  \mu_4 = M_{\overline{v}} M_{\overline{\sigma}}  \\
  \end{align}
\end{subequations}
where
\begin{subequations}
  \begin{align}
    M_{\underline{v}} = \frac{v - \underline{v}}{\overline{v} - \underline{v}} \ , \ M_{\overline{v}} = 1 - M_{\underline{v}} \\
    M_{\underline{\sigma}} = \frac{\sigma - \underline{\sigma}}{\overline{\sigma} - \underline{\sigma}} \ , \ M_{\overline{\sigma}} = 1 - M_{\underline{\sigma}}
  \end{align}
\end{subequations}

\noindent The chosen control scheme for this dynamic loop has the following expression
\begin{equation}
    \bm{u_f} = \bm{K_D} \bm{x_D} + \bm{N}_{ff} \bm{r_D}
\end{equation}

\noindent where $\bm{K_D}$ is the controller computed at every iteration by \eqref{eq:interpolation} (see Figure \ref{fig:control_scheme}), $\bm{N_{ff}}$ is a feedforward matrix, $\bm{x_D}$ is the dynamic state vector \eqref{eq:state_vector_LPV}, $\bm{r_D}$ represents the reference vector which corresponds with the kinematic control signal $\bm{u_C}$, and $\bm{u_f}$ is control input to the filter added \eqref{eq:filter}. At this point, the dynamic control action $\bm{u_D}$ that is applied over the vehicle will be the output of applying $\bm{u_f}$ to the filter. \\

\noindent The feedforward matrix has been computed following the next expression
\begin{equation}
    \bm{N_{ff}} = \left[ \bm{C} \left( -\bm{B} \bm{K} - \bm{A} \right)^{-1} \bm{B} \right]^{-1}
\end{equation}

\noindent where matrices $\bm{A}$ and $\bm{B}$ are the ones presented in \eqref{eq:dynamic_state_space_vectors_augmented} (i.e., without considering the added integrator which cause the matrix $\bm{A}$ cannot be inverted), $\bm{K}$ is a sub-block of $\bm{K}_D$ in which the last column has been omitted as it is proposed in \cite{franklin1998digital}.
Matrix $\bm{C}$ is of the form

 \begin{equation}
        \bm{C} = \left[\begin{array}{ccccc}
      	  1 & 0 & 0 & 0 & 0 \\
     	  0 & 0 & 1 & 0 & 0 \\
      \end{array}\right]
\end{equation}

\noindent It is interesting to note that, for controlling the vehicle in the interval $v_d \in \left[ 0, 1 \right]$ a translation has been applied.
Thus, this means that when computing the controller at $v_d=0\frac{m}{s}$, we are actually computing the controller at $v_d=1\frac{m}{s}$ and using it as we are in $v_d=0\frac{m}{s}$. In this way, we avoid to develop a hybrid control for this reduced velocity interval. \\

\begin{table*}[ht]
    \caption{Dynamic controllers $\bm{K_{i}}$  for each one of the vertexes $i$ of the polytope shown in Figure \ref{fig:polytopic_dynamic}.}
    \centering
    \begin{tabular}{*{2}{c}}
        \hline
        \\
        $\bm{K_1} = 10^{4}\left[\begin{array}{cccccc}
        -0.7845 & -0.1760 & -0.0802 & -0.0002 & 0.0027 & -0.3280 \\
        0.0000 & 0.1073 & -0.2441 & 0.0000 & -0.0107 & 0.5575
        \end{array}\right]$ \\
        \\
        \\
        $\bm{K_2} = 10^{4} \left[\begin{array}{cccccc}
        -0.6129 & -0.7835 & -0.2095 & -0.0000 & 0.0010 & -1.3048 \\
        0.0003 & 0.1559 & -0.2622 & 0.0000 & -0.0111 & 0.6480
        \end{array}\right]$ \\
        \\
        \\
        $\bm{K_3} = 10^{4} \left[\begin{array}{cccccc}
       -1.7823 & -0.1366 & -0.1164 & 0.0001 & 0.0046 & -0.3888 \\
        0.0003 & 0.0988 & -0.2684 & 0.0000 & -0.0111 & 0.5564
        \end{array}\right]$ \\
        \\
        \\
        $\bm{K_4} = 10^{4} \left[\begin{array}{cccccc}
       -0.6104 & -0.4180 & -0.2686 & -0.0000 & 0.0044 & -3.1728 \\
        0.0002 & 0.1591 & -0.2621 & 0.0000 & -0.0111 & 0.6489
        \end{array}\right]$ \\
        \\
        \hline
    \end{tabular}
    \label{table:dynamic_controllers}
\end{table*}

\subsection{Kinematic LPV control design}

Kinematic control is in charge of controlling the position, orientation and linear velocity by means of actuating over the linear and angular velocities of the vehicle. \\

\noindent At this moment, the kinematic LPV model \eqref{eq:kinematic_LPV_matrices} is employed for solving the Problem \ref{problem1}. Three scheduling variables ($v_d, \omega$ and $\theta_e$) are bounded in the following intervals
\begin{equation*}
    v_d \in [1, 18]  \  \frac{m}{s} \ \ , \ \ \ \ \omega \in [-1.417, 1.417]  \  \frac{rad}{s} 
\end{equation*}
\begin{equation*}
    \theta_{e} \in [-0.139, 0.139]  \  rad
\end{equation*}

\noindent The control design matrices $\bm{Q}$ and $\bm{R}$ are presented in Table \ref{table:RMSEs} and parameter $\gamma$ is set as $0.01$.
The problem \ref{problem1} returns for this kinematic case the matrices $\bm{P} \in \mathbb{R}^3$, $\bm{Y} \in \mathbb{R}^2$ and $\bm{W} \in \mathbb{R}^{2 \times 3}$.
Then, by inserting $\bm{P}$ and $\bm{Q}$ in \eqref{eq:controller_obtaining} the vertex controllers $\bm{K_i}$ are obtained (see Table \ref{table:kinematic_controllers}). \\

\noindent It is important to remark that the Problem \ref{problem1} has a different configuration with respect to the dynamic case. 
The first LMI of \eqref{eq:LQR-LMI-introduction} is negative and an additional LMI has been added to the Problem \ref{problem1}
\begin{equation}
\label{eq:new-LMI}
    (\bm{A}_i\bm{P} + \bm{B}\bm{W}_i) + (\bm{A}_i \bm{P} + \bm{B}\bm{W}_i)^T + 2 \beta \bm{P} < 0
\end{equation}

\begin{figure}[H]
  \centering
  \includegraphics[width=60mm]{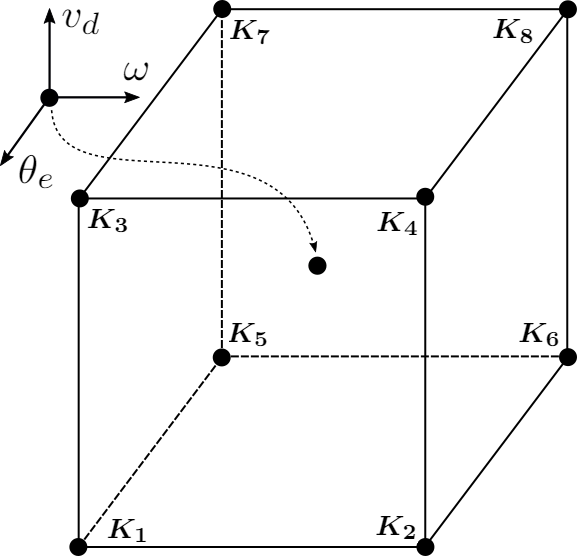}
  \caption{Kinematic polytopic region defined by three scheduling variables.}
  \label{fig:polytopic_kinematic}
\end{figure}

\noindent Being $\beta = 0$, the LMI establishes a threshold for ensuring only stability. Thus, in order to increase the kinematic loop performance $\beta$ can be raised being always positive. Observe that, due to the number of scheduling variables, the polytopic region is defined as a three-dimensional cube (see Figure \ref{fig:polytopic_kinematic}).

\noindent In this kinematic case, the weighting variables $\mu_i$ are computed as
\begin{subequations}
    \begin{align}
    \mu_1 = M_{\underline{v_d}} M_{\underline{\omega}} M_{\underline{\theta_e}} \ , \ \ \mu_5 = M_{\overline{v_d}} M_{\underline{\omega}} M_{\underline{\theta_e}}  \\
    \mu_2 = M_{\underline{v_d}} M_{\underline{\omega}} M_{\overline{\theta_e}}  \ , \ \ \mu_6 = M_{\overline{v_d}} M_{\underline{\omega}} M_{\overline{\theta_e}}  \\
    \mu_3 = M_{\underline{v_d}} M_{\overline{\omega}} M_{\underline{\theta_e}}  \ , \ \ \mu_7 = M_{\overline{v_d}} M_{\overline{\omega}} M_{\underline{\theta_e}}  \\
    \mu_4 = M_{\underline{v_d}} M_{\overline{\omega}} M_{\overline{\theta_e}}  \ , \ \ \mu_8 = M_{\overline{v_d}} M_{\overline{\omega}} M_{\overline{\theta_e}} 
    \end{align}
\end{subequations}
where
\begin{subequations}
  \begin{align}
    M_{\underline{v_d}} = \frac{v_d - \underline{v_d}}{\overline{v_d} - \underline{v_d}} \ , \ M_{\overline{v_d}} = 1 - M_{\underline{v_d}} \\
    M_{\underline{\omega}} = \frac{\omega - \underline{\omega}}{\overline{\omega} - \underline{\omega}} \ , \ M_{\overline{\omega}} = 1 - M_{\underline{\omega}} \\
    M_{\underline{\theta_e}} = \frac{\theta_e - \underline{\theta_e}}{\overline{\theta_e} - \underline{\theta_e}} \ , \ M_{\overline{\theta_e}} = 1 - M_{\underline{\theta_e}}
  \end{align}
\end{subequations}

\noindent The following state feedback control law has been used for controlling the kinematic behaviour loop

\begin{equation}
    \bm{u_C} = \bm{K_C} \bm{x_C} + \bm{r_C}
\end{equation}

\noindent where $\bm{K}_C$ represents the interpolated kinematic controller obtained with \eqref{eq:interpolation}, $\bm{x}_C$ represents the kinematic state vector \eqref{eq:kinematic_state_space_vectors} and $\bm{r}_C$ is the reference. Such a reference is provided by a trajectory planner (see Figure \ref{fig:control_scheme}). \\

\begin{table*}[ht]
    \caption{Kinematic controllers $\bm{K_i}$ for each one of the vertexes $i$ of the polytope shown in Figure \ref{fig:polytopic_kinematic}.}
    \centering
    \begin{tabular}{*{2}{c}}
        \hline
        \\
        $\bm{K_1} = \left[\begin{array}{cccccc}
            0.7099 & 0.5078  & -0.0238 \\
            0.1899 & 0.3083  & 1.5405
        \end{array}\right]$ , \
        &
        $\bm{K_2} = \left[\begin{array}{cccccc}
            0.7099 & 0.5078  & -0.0238 \\
            0.1899 & 0.3083  & 1.5405
        \end{array}\right]$
        \\
        \\      
        $\bm{K_3} = \left[\begin{array}{cccccc}
      	    0.7373  &  0.2156 & 0.0158 \\
            0.1792  &  2.0131  &  4.0841
        \end{array}\right]$  , \ 
        &
        $\bm{K_4} = \left[\begin{array}{cccccc}
      	    0.7373  &  0.2156 & 0.0158 \\
            0.1792  &  2.0131  &  4.0841
        \end{array}\right]$ \\        
        \\
        \\
        $\bm{K_5} = \left[\begin{array}{cccccc}
            0.7099 & -0.5078  & 0.0238 \\
            -0.1899 & 0.3083  & 1.5405
        \end{array}\right]$  , \
        &
        $\bm{K_6} = \left[\begin{array}{cccccc}
            0.7099 & -0.5078  & 0.0238 \\
            -0.1899 & 0.3083  & 1.5405
        \end{array}\right]$  \\        
        \\
        \\
        $\bm{K_7} = \left[\begin{array}{cccccc}
      	    0.7373  &  -0.2156 & -0.0158 \\
            -0.1792  &  2.0131  &  4.0841
        \end{array}\right]$  , \
        &
        $\bm{K_8} = \left[\begin{array}{cccccc}
      	    0.7373  &  -0.2156 & -0.0158 \\
            -0.1792  &  2.0131  &  4.0841
        \end{array}\right]$ \\        
     
        \\
        \hline
    \end{tabular}
    \label{table:kinematic_controllers}
\end{table*}


\section{Simulation Results}
\label{sec:results}

\begin{figure}[H]
    \centering
    \includegraphics[scale=0.6]{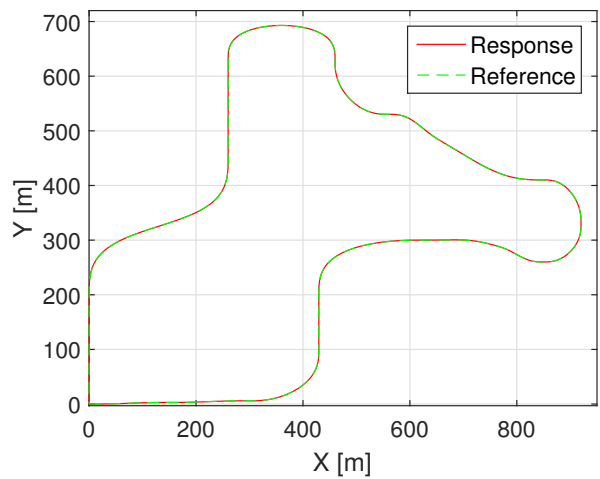}
    \caption{Proposed circuit for simulation and the result of solving the mixed control problem.}
    \label{fig:dynamic_LPV_results_1}
\end{figure}

The simulation scenario chosen for testing the automatic control strategy tries to cover different driving situations as acceleration stage, velocity reduction on curves and slow down at the end of the circuit.
Such scenario is characterized by having curves of different geometry and curvature with the idea of testing the autonomous guidance system at several behaviours (see Figure \ref{fig:dynamic_LPV_results_1}).

\noindent Considering this information (circuit shape and varying velocity), a trajectory planner is in charge on generating a feasible trajectory by means of using a polynomial curve generation method \cite{GuarinoPiazziRomano}.
This consists on computing continuous and differentiable curves (velocities and accelerations) under an overall constrained vehicle acceleration.
Thus, in an offline mode, this planner algorithm generates the linear and angular velocity references as well as desired positions and orientations for the outer control loop (i.e., kinematic control). \\

\noindent The adjustment of the LPV-LQR parameters ($\bm{Q}$, $\bm{R}$ and $\gamma$) is made by means of using the root mean square error (RMSE) approach.
This measure allows to find suitable control parameters by minimizing it.
Linear velocity, angular velocity and lateral errors are chosen by an exhaustive search. Moreover, $\eta$ and $\beta$ have been chosen with the aim of increasing the performance of the closed loop system.
Table \ref{table:RMSEs} shows some RMSE results for different control adjustments and the one chosen in the simulations.

\begin{table*}[ht]
        \caption{ Root mean square error obtained for three different configurations of the LQR controllers. The values of $\bm{Q}$ and $\bm{R}$ represent the diagonal values of each matrix. The bold raw is the configuration chosen for the performed simulation.}
        \centering
        \begin{tabular}{|*{7}{c|}}
        \hline
        \multicolumn{3}{|c }{RMSE} & \multicolumn{2}{|c}{Kinematic control design} & \multicolumn{2}{|c|}{Dynamic control design}   \\
        \hline
        \multicolumn{1}{|c}{V} & \multicolumn{1}{|c}{ $\omega$} & \multicolumn{1}{|c}{ Y} & \multicolumn{1}{|c}{$\bm{Q}$} & \multicolumn{1}{|c}{$\bm{R}$} & \multicolumn{1}{|c}{$\bm{Q}$} & \multicolumn{1}{|c|}{$\bm{R}$}   \\
        \hline
        \hline
        0.121  & 0.035 & 0.0177  & [1, 1, 1] & [0.004, 0.0001] & [0.01, 0.01, 0.01, 0.01, 10, 3000] & [0.005, 0.6]  \\
        \hline
        0.124  & 0.031 & 0.0196  & [3, 5, 15] & [0.04, 0.01] & [0.01, 0.01, 0.01, 0.01, 100, 30000] & [0.005, 0.6]  \\
        \hline
        0.076  & 0.0127 & 0.0213  & [10, 3, 15] & [0.4, 0.001] & [0.01, 0.01, 0.01, 0.01, 1000, 90000] & [0.005, 0.6] \\
        \hline
        $\bm{0.045}$  & $\bm{0.0077}$ & $\bm{0.05}$  & $\bm{[3, 2, 20]}$ & $\bm{[0.5, 0.001]}$ & $\bm{[0.01, 0.01, 0.01, 0.01, 100000, 90000]}$ & $\bm{[0.01, 10]}$  \\
        \hline
        \end{tabular}
        \label{table:RMSEs}
\end{table*}

\vspace{4mm} 

\noindent In the tuning process, we have observed that the vehicle lateral behaviour is more difficult to control due to the changing reference. Hence, the weight in $\bm{Q}_D$ corresponding to the dynamic integral state has been set much bigger than the rest. The same occurs in matrix $\bm{R}_D$. \\

\noindent The sample times used in both control loops are 0.1 s and 0.01 s for kinematic and dynamic loops, respectively.
The control strategy jointly with the trajectory planner are tested in Matlab environment.

\begin{figure}[H]
    \centering
    \subfigure[]{\includegraphics[scale=0.6]{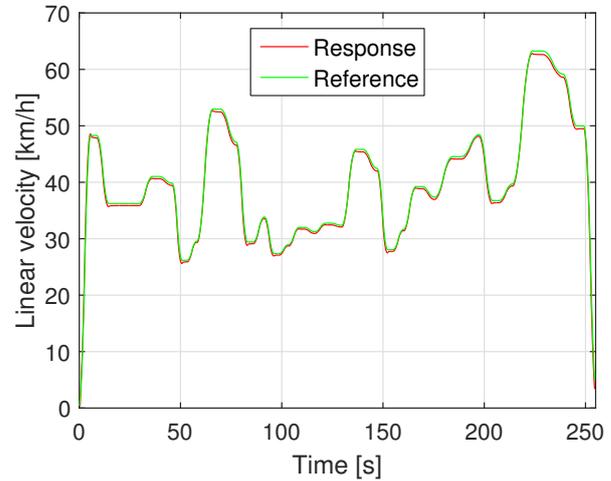}} \\
    \subfigure[]{\includegraphics[scale=0.6]{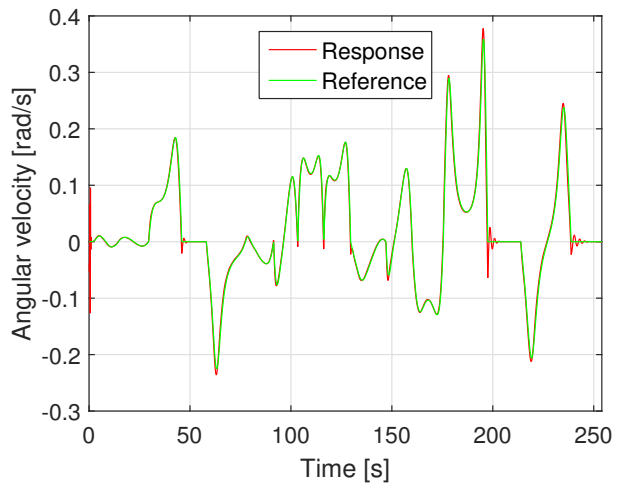}}
    \caption{Results of applying the kinematic-dynamic LPV control for solving the mixed problem. a) linear velocity reference and response. b) desired and simulated angular velocities.}
    \label{fig:dynamic_LPV_results_2}
\end{figure}

\begin{figure}[H]
    \centering
    \subfigure[]{\includegraphics[scale=0.6]{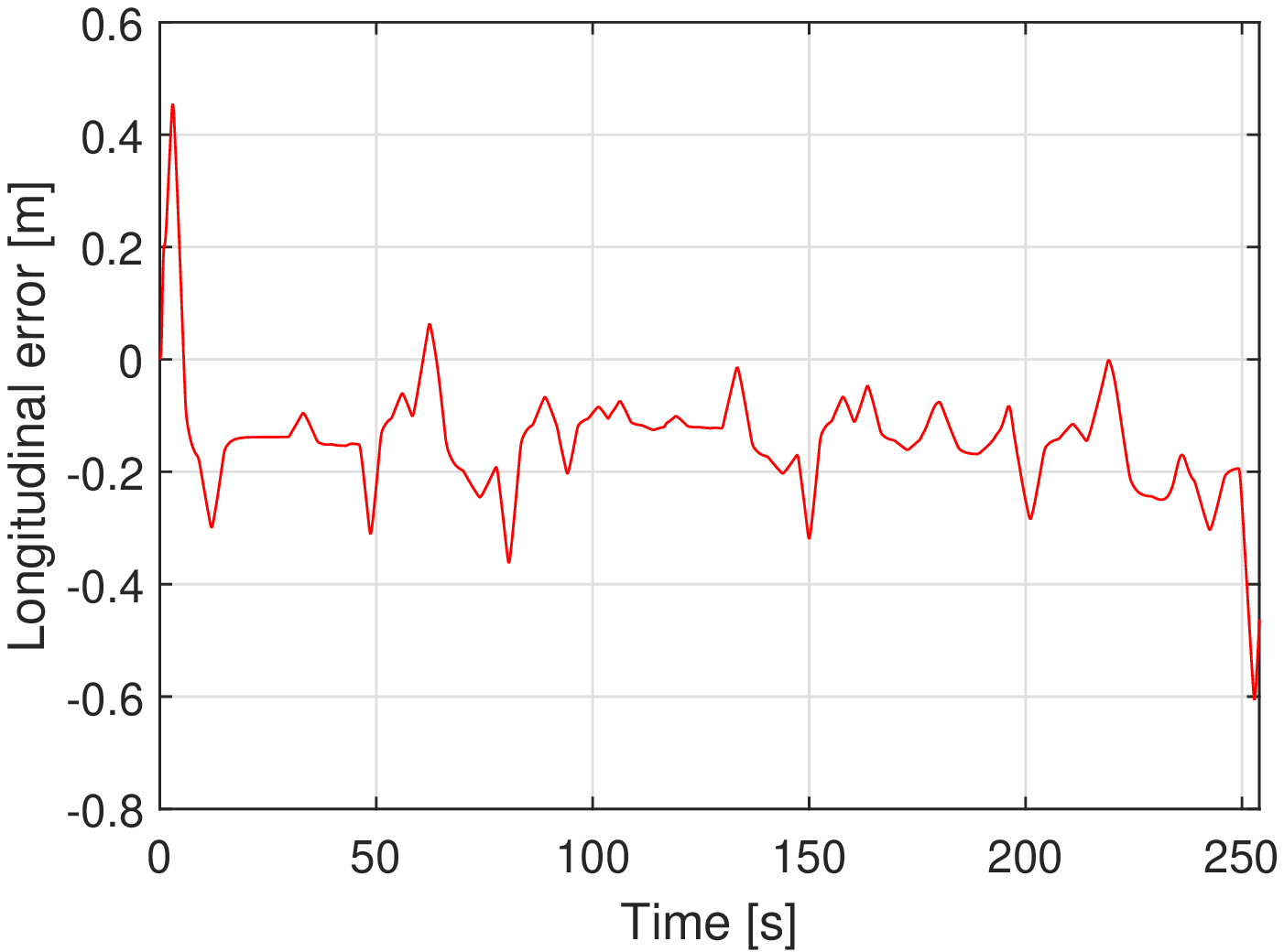}} \\
    \subfigure[]{\includegraphics[scale=0.6]{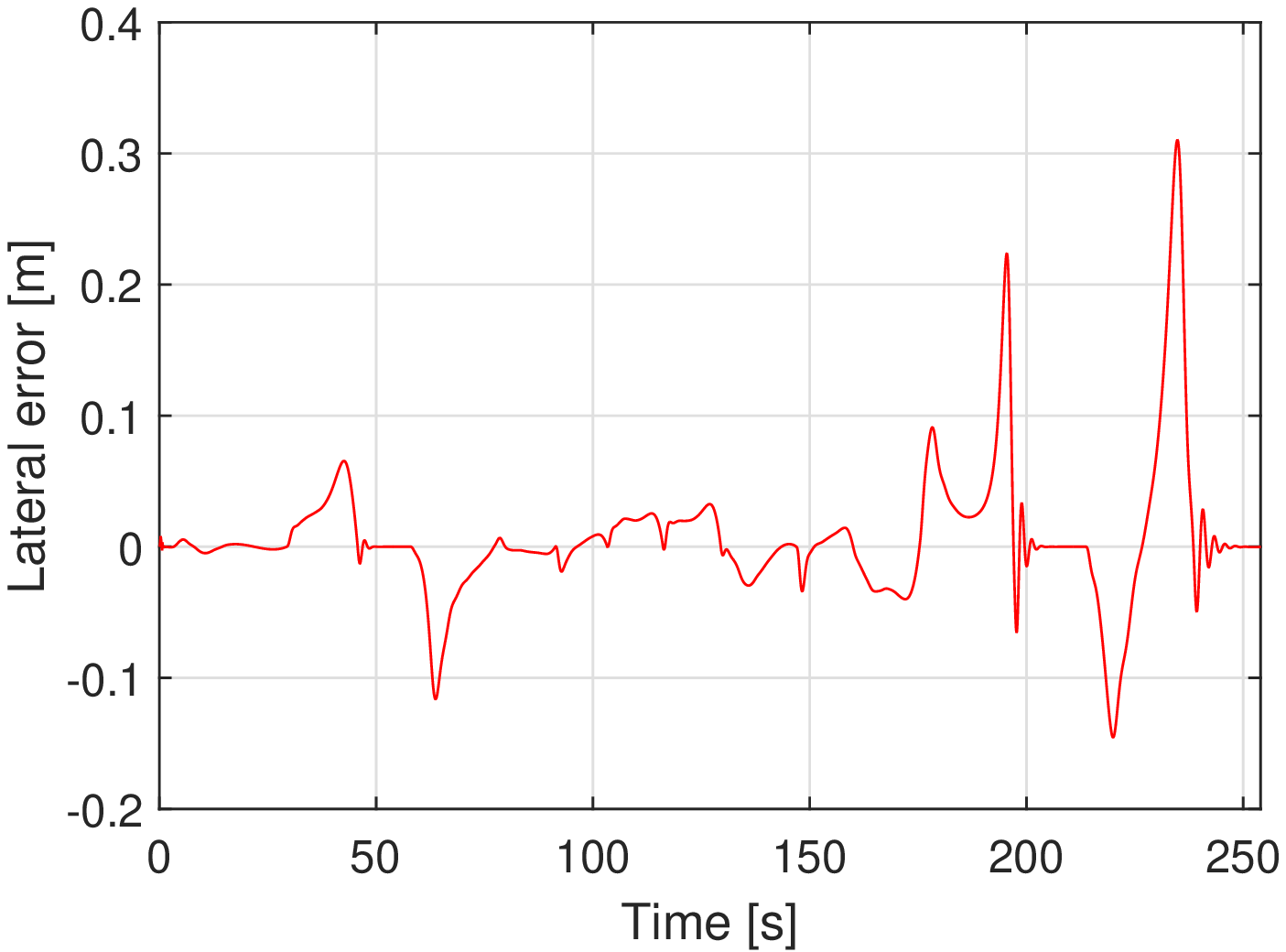}}
    \caption{Position errors. a) vehicle longitudinal error along the circuit. b) vehicle lateral error. }
    \label{fig:dynamic_LPV_results_3}
\end{figure}

\begin{figure}[H]
    \centering
    \subfigure[]{\includegraphics[scale=0.6]{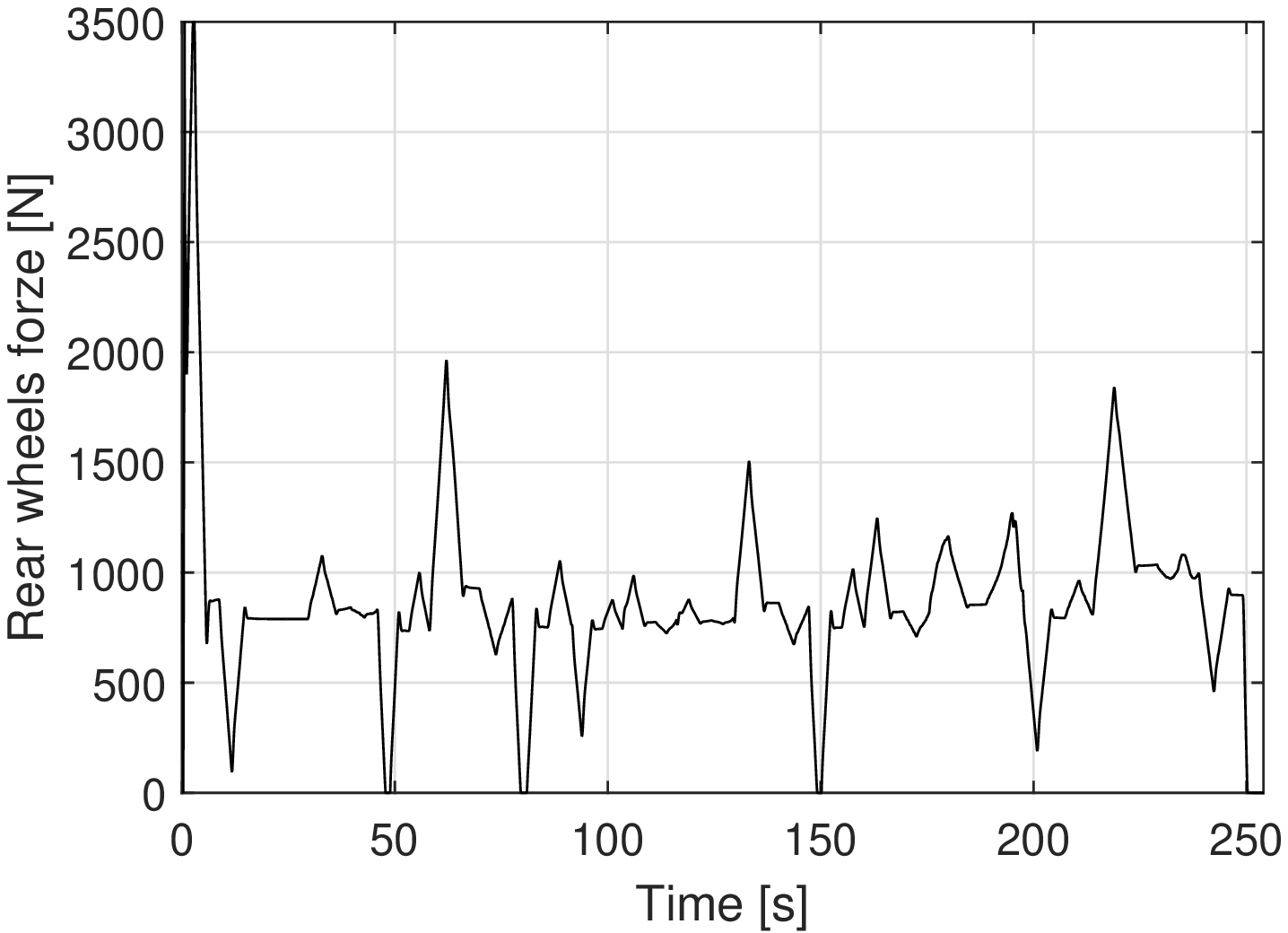}} \\
    \subfigure[]{\includegraphics[scale=0.6]{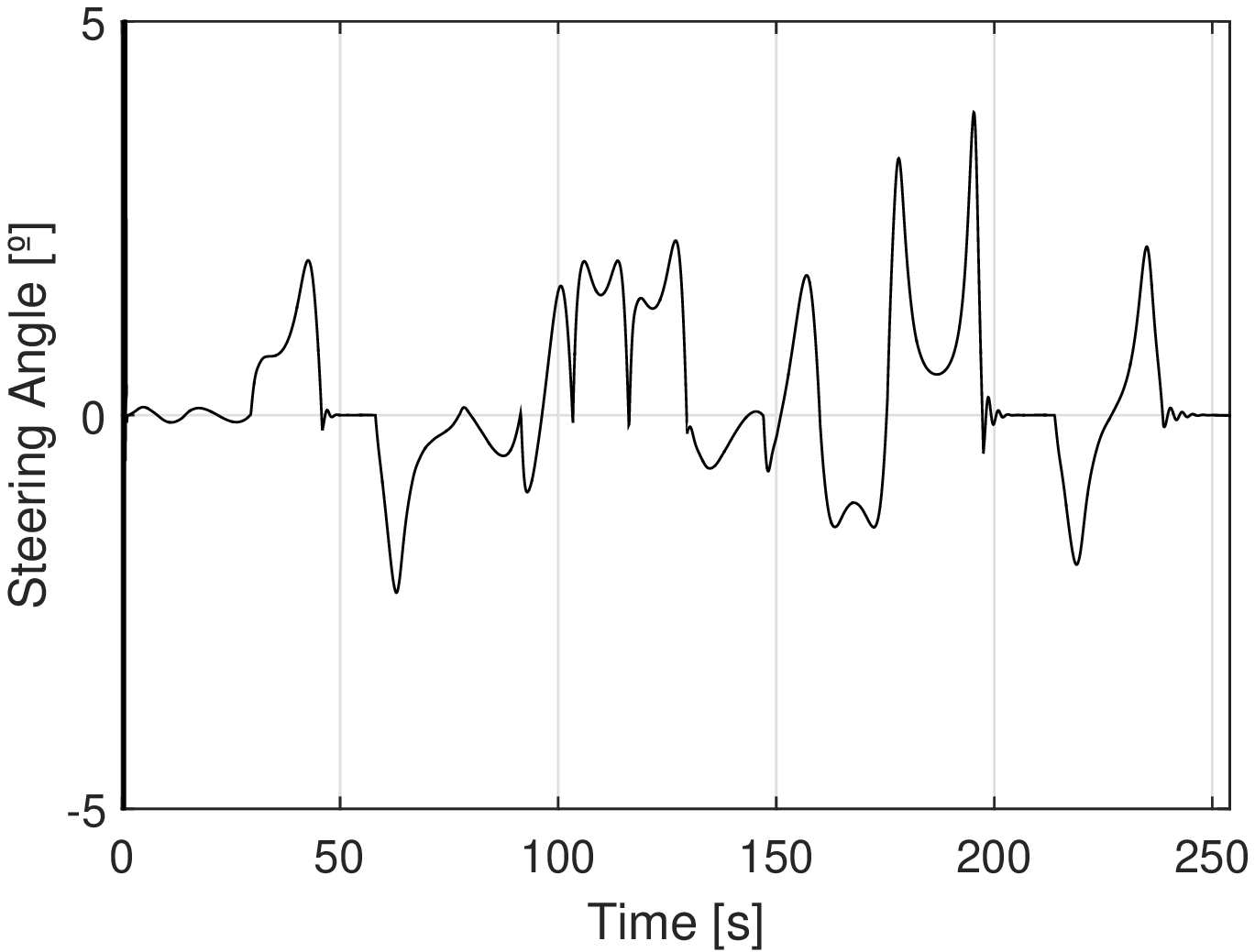}}
    \caption{Resulting control actions of applying the kinematic-dynamic LPV control for solving the mixed problem.}
    \label{fig:dynamic_LPV_results_4}
\end{figure}

\begin{figure}[!htbp]
    \centering
    \includegraphics[scale=0.6]{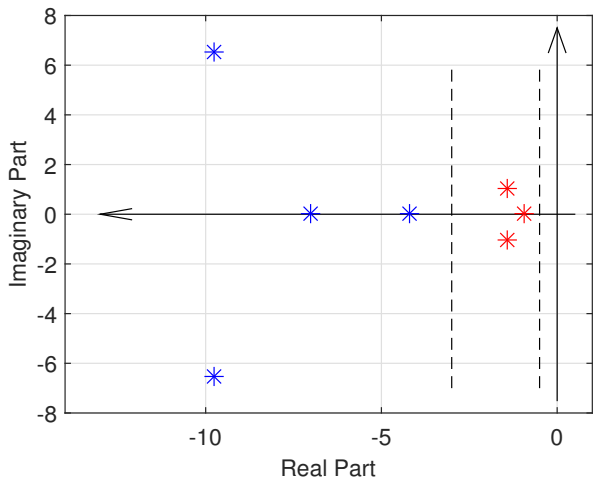}
    \caption{Pole locus of the system in a particular operating point ($v=8.33 \frac{m}{s}$ and $\omega=0.05 \frac{rad}{s}$). Blue marks are the five slower poles of the dynamic loop and the red ones are the kinematic poles. Vertical dashed lines represent the hyperplanes $\eta=3$ and $\beta=0.5$. }
    \label{fig:rlocus}
\end{figure}

\noindent Figures \ref{fig:dynamic_LPV_results_2}-\ref{fig:dynamic_LPV_results_4} show the results of the vehicle in the simulated circuit. Finally, Figure \ref{fig:rlocus} represents the location of the closed loop poles of kinematic and dynamic controllers, and the thresholds for the decay rate ($\eta$ and $\beta$) used in their design. \\

\noindent Figure \ref{fig:dynamic_LPV_results_2}.a) shows that in the linear velocity response, the controller cannot eliminate the error completely due to curves in the path make difficult to maintain null error in both, angular and linear velocities. However, the maximum error achieved is $0.5 \frac{km}{h}$.
In Figure \ref{fig:dynamic_LPV_results_2}.b), it can be seen how, for the angular velocity, the integral action makes the controller to perform better than the linear velocity with respect to the reference.
Even so, the angular response presents some overshoot behaviour at some time instants.
The controller adjustment may be one of the reasons, but the main reason is the high abruptness of the angular velocity reference at the end of the curves producing a rough behaviour on the vehicle. \\

\noindent Figure \ref{fig:dynamic_LPV_results_3} depicts the position errors. The mitigation of these errors is crucial for achieving a good autonomous guidance. 
However, an almost null lateral error is most important due to this ensures that the vehicle is driving over the road center. In our results, longitudinal error is no longer than $0.4 m$ in normal driving (i.e. neither accelerating nor braking).
Lateral error remains in the scale of few decimeters being increased when both velocities (angular and linear) increase. \\

\noindent Figure \ref{fig:dynamic_LPV_results_4} shows the resulting control actions. They are appropriate signals, specially the steering one which shows an smooth behaviour. The force applied to the wheels seems to be a bit sharp when the vehicle arrives at curves.
It should be emphasized that when force control action goes to zero means the controller wants to brake the vehicle.
However, in this work the brake system has not been addressed and therefore the only way to slow down the vehicle is applying null force.
Note also that the steering angle signal in the first little part of the simulation is so abrupt. This behaviour is due to longitudinal and angular behaviours are highly coupled and the starting stage deals with high linear accelerations.

Figure \ref{fig:rlocus} illustrates the closed poles for the kinematic and dynamic loops at a given operating point.
It can be observed that the poles of both loops satisfy the constraints imposed by the corresponding decay rates $\eta$ and $\beta$ (see \eqref{eq:new-LMI} and \eqref{eq:LQR-LMI-introduction}). The satisfaction of this condition allows to design both loops separately, since the dynamic control presents a faster dynamic behaviour than the kinematic one.

\section{Conclusions}
\label{sec:conclusions}
In this work, a gain scheduling LPV control scheme has been introduced for solving the mixed lateral and logitudinal control problem of autonomous vehicle guidance.
For such a purpose, a cascade methodology has been employed for controlling an outer loop (kinematic vehicle behaviour) and an inner loop (dynamic vehicle behaviour).
For the design of both controllers, the use of two separated models describing different vehicle behaviours (kinematic and dynamic models) allows to solve two smaller LMI-LQR problems instead of a larger and more complex one presenting very different dynamic behaviours.
Moreover, an approach based on cascade design of the kinematic and dynamic controllers has been adopted with the aim of increasing the performance of the system. It consists on forcing that the inner closed loop dynamics behaves faster than the outer closed loop one.
The obtained gain scheduling LPV control approach, jointly with a trajectory generation module, has presented suitable results in a simulated city scenario.
As a future work, this mixed control approach will be tested in a real city situation and the trajectory planner will be improved for providing smoother references.


\section*{Acknolwedgment}
This work has been funded by the Spanish Ministry of Economy and Competitiveness (MINECO) and FEDER through the projects ECOCIS (ref. DPI201348243. C2-1-R) and HARCRICS (ref. DPI2014-58104-R). The author is supported by a FI AGAUR grant (ref 2017 FI B00433).

\bibliography{mybibfile}

\begin{thebibliography}{10}
\expandafter\ifx\csname url\endcsname\relax
  \def\url#1{\texttt{#1}}\fi
\expandafter\ifx\csname urlprefix\endcsname\relax\def\urlprefix{URL }\fi
\expandafter\ifx\csname href\endcsname\relax
  \def\href#1#2{#2} \def\path#1{#1}\fi

\bibitem{EuroParl}
L.~Van~Woensel, G.~Archer, L.~Panades-Estruch, D.~Vrscaj, Ten technologies
  which could change our lives, European Union: Brussels, Belgium.

\bibitem{WorldUrmanizationProspects}
U.~Nations,
  \href{$https://esa.un.org/unpd/wup/publications/files/wup2014-highlights.Pdf$}{{World
  Urbanization Prospects. The 2014 Revision}}, United Nations, Department of
  Economic and Social Affairs, Population Division (2014) (2014).
\newline\urlprefix\url{$https://esa.un.org/unpd/wup/publications/files/wup2014-highlights.Pdf$}

\bibitem{SAE}
P.~WARRENDALE,
  \href{$https://www.sae.org/misc/pdfs/automated_driving.pdf$}{{Levels of
  automation for on-road vehicles}}, Society of Automotive Engineers (SAE)
  (2014).
\newline\urlprefix\url{$https://www.sae.org/misc/pdfs/automated_driving.pdf$}

\bibitem{H_infinity}
N.~Nawash, H-infinity control of an autonomous mobile robot, Master of Science
  Thesis, Cleveland State University.

\bibitem{alcala2016comparison}
E.~Alcal{\'a}, L.~Sellart, V.~Puig, et~al, Comparison of two non-linear
  model-based control strategies for autonomous vehicles, in: Control and
  Automation (MED), 2016 24th Mediterranean Conference on, IEEE, 2016, pp.
  846--851.

\bibitem{indiveri1999kinematic}
G.~Indiveri, Kinematic time-invariant control of a 2d nonholonomic vehicle, in:
  Decision and Control, 1999. Proceedings of the 38th IEEE Conference on,
  Vol.~3, IEEE, 1999, pp. 2112--2117.

\bibitem{blavzivc2010takagi}
S.~Bla{\v{z}}i{\v{c}}, Takagi-sugeno vs. lyapunov-based tracking control for a
  wheeled mobile robot, WSEAS Transactions on Systems and Control 5~(8) (2010)
  667--676.

\bibitem{LyapunovBasedControl}
W.~E.Dixon, D.~M.Dawson, E.~Zergeroglu, A.~Behal, Nonlinear Control of Wheeled
  Mobile Robots, Springer, 2000.

\bibitem{nemeth2016lpv}
B.~N{\'e}meth, P.~G{\'a}sp{\'a}r, J.~Bokor, Lpv-based integrated vehicle
  control design considering the nonlinear characteristics of the tire, in:
  American Control Conference (ACC), 2016, IEEE, 2016, pp. 6893--6898.

\bibitem{LinearMPC}
C.~Olsson, Model complexity and coupling of longitudinal and lateral control in
  autonomous vehicles using model predictive control (2015).

\bibitem{gonzalez2011robust}
R.~Gonz{\'a}lez, M.~Fiacchini, J.~L. Guzm{\'a}n, T.~{\'A}lamo,
  F.~Rodr{\'\i}guez, Robust tube-based predictive control for mobile robots in
  off-road conditions, Robotics and Autonomous Systems 59~(10) (2011) 711--726.

\bibitem{farrokhsiar2013integrated}
M.~Farrokhsiar, G.~Pavlik, H.~Najjaran, An integrated robust probing motion
  planning and control scheme: A tube-based mpc approach, Robotics and
  Autonomous Systems 61~(12) (2013) 1379--1391.

\bibitem{NLMPC}
Y.~Gao, A.~Gray, H.~E. Tseng, F.~Borrelli, A tube-based robust nonlinear
  predictive control approach to semi-autonomous ground vehicles, Vehicle
  System Dynamics: International Journal of Vehicle Mechanics and Mobility.

\bibitem{JournalMPC}
A.~Carvalho, S.~Lefévre, G.~Schildbach, J.~Kong, F.~Borrelli, Automated
  driving: The role of forecasts and uncertainty — a control perspective,
  European Journal of Control.

\bibitem{apkarian1995self}
P.~Apkarian, P.~Gahinet, G.~Becker, Self-scheduled h infinite control of linear
  parameter-varying systems: a design example, Automatica 31~(9) (1995)
  1251--1261.

\bibitem{GuarinoPiazziRomano}
C.~G.~L. Bianco, A.~Piazzi, M.~Romano, Velocity planning for autonomous
  vehicles, IEEE Intelligent Vehicles Symposium.

\bibitem{duan2013lmis}
G.-R. Duan, H.-H. Yu, LMIs in control systems: analysis, design and
  applications, CRC press, 2013.

\bibitem{floater2006general}
M.~S. Floater, K.~Hormann, G.~K{\'o}s, A general construction of barycentric
  coordinates over convex polygons, advances in computational mathematics
  24~(1) (2006) 311--331.

\bibitem{franklin1998digital}
G.~F. Franklin, J.~D. Powell, M.~L. Workman, Digital control of dynamic
  systems, Vol.~3, Addison-wesley Menlo Park, CA, 1998.

\end{thebibliography}

\end{document}